# Imperfect chirality at exceptional points in optical whispering-gallery microcavities

Junda Zhu[1,2], Changqing Wang[3], Can Tao[4], Zhoutian Fu[3], Haitao Liu[4,*], Fang Bo[1,†], Lan Yang[3,‡], Guoquan Zhang[1,§], and Jingjun Xu[1,¶]

[1]*The MOE Key Laboratory of Weak Light Nonlinear Photonics, TEDA Institute of Applied Physics and School of Physics, Nankai University, Tianjin 300457, China*
[2]*College of Physics and Materials Science, Tianjin Normal University, Tianjin 300387, China*
[3]*Department of Electrical and Systems Engineering, Washington University, St Louis, MO, USA*
[4]*Tianjin Key Laboratory of Micro-scale Optical Information Science and Technology, Institute of Modern Optics, College of Electronic Information and Optical Engineering, Nankai University, Tianjin 300350, China*

Non-Hermitian systems have attracted considerable attention for their broad impacts on various physical platforms and peculiar applications. In non-Hermitian systems, both eigenvalues and eigenstates simultaneously coalesce at exceptional points (EPs). As one of the remarkable features of EPs, the field chirality is commonly considered perfect, which is utilized as an intriguing feature to control wave propagation and regarded as a criterion of EP. However, in this work, we discover an imperfect chirality of eigenmodes at the EPs in an optical whispering gallery mode (WGM) microcavity perturbed by two strong nanoscatterers. This counterintuitive phenomenon originates from a strong frequency-dependence of the scattering between the counterpropagating waves at an "effective scatterer", which could be explained by a first-principle-based model considering a dynamic multiple-scattering process of the azimuthally propagating modes. We find that the generally imperfect chirality at the EP tends to be globally perfect with the decrease of the scattering effect induced by the nanoscatterers. Furthermore, the chirality also becomes locally perfect with the decrease of the relative azimuthal angle between the two strong nanoscatterers. This work provides a new understanding of the general properties of chirality at EPs. It will benefit the potential applications enabled by the chirality features of non-Hermitian systems at EPs.

Exceptional points (EPs) [1,2] are spectral degeneracies in the parameter space of a non-Hermitian system, at which the eigenstates and their associated eigenvalues coalesce. EPs have been studied in various physical platforms ranging from atom systems [3], atom-cavity systems [4], acoustic systems [5], and optical systems [6] to optomechanical systems [7]. In optical systems, unique features at EPs have been utilized for enhanced sensing [8-10], laser linewidth broadening [11], asymmetric mode switching [12-14], etc.

Chirality is one of the unique feature at EP, and in general means a specific phase relation between two dominant states in a non-Hermitian system [15,16]. For an optical whispering-gallery-mode (WGM) microcavity system, chirality means a resonant/eigenmode with a dominant clockwise (CW) or counterclockwise (CCW) rotation [17]. By introducing parity-time symmetric refractive index modulation [18], judiciously deforming the shape of the microcavity [17] or tuning light scattering within the mode volume of the microcavity [19,20], one can steer the system to an EP and achieve degenerate eigenmodes with strong spatial chirality. Owing to the chirality at the EP, numerous interesting phenomena and applications have been demonstrated, such as orbital angular momentum lasing [18], chiral lasing [20], chiral absorbing [21,22], and electromagnetically induced transparency [23,24].

Previous works predict that the chirality is perfect at the EP in an optical WGM microcavity perturbed by two nanoscatterers with an effective non-Hermitian Hamiltonian approach [19]. A perfect chirality implies that there only exists a pure CW or CCW WGM component in the resonant modes, and is regarded as a criterion of the EP in the experiment [8,20,25]. However, this letter will demonstrate that the chirality of the resonant modes at the EP is generally imperfect and a perfect chirality only exists conditionally.

Here we consider a two-dimensional cylindrical microcavity [26] (with radius $R_0$) perturbed by two nanoholes [27] inside the microcavity to induce a strong scattering effect (see SM, Sec. 1 [28]). As illustrated in Fig. 1(a), two sectorial-shaped [26] nanoholes with different radial lengths $r_j$ ($j$=1,2), azimuthal ranges $\theta_j$, and distances $d_j$ away from the cavity boundary are considered.

*Definition of the local chirality*. —To define the local chirality of the electromagnetic field in different azimuthal regions divided by the two nanoholes, we introduce azimuthally propagating modes (APMs) [26,29] and "effective scatterer" [26].

(i) APM is a waveguide mode with an electromagnetic field defined on the cross-section of $\phi$=constant, and thus can be used as a local basis. Differently, WGM, the basis adopted in previous works [19-21,23], is a resonant mode (i.e. the quasinormal mode [30,31] with discrete complex eigen/resonance frequencies) in an unperturbed microcavity, and thus is defined over the whole azimuthal range of [$0,2\pi$]. Therefore, the WGM acts as a global basis and cannot describe the local chirality. Here we consider the two counterpropagating APMs [26,29] that form the pair of degenerate WGMs under the resonance condition [32].

(ii) A single "effective scatterer" is defined as the two nanoholes along with the azimuthal region between them, as illustrated by the blue region in Fig. 1(a) [26]. Then the electromagnetic field of the resonant mode can be expressed as the superposition of the CCW and CW APMs (as the local basis),

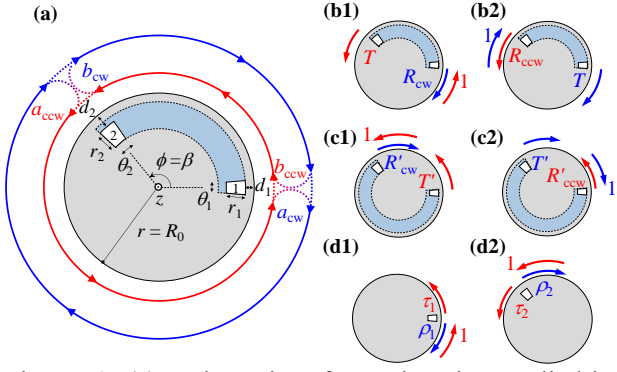

Figure 1 (a) Schematic of a $z$-invariant cylindrical microcavity with two nanoholes. $r$ and $\phi$ denote the radial and azimuthal coordinates, respectively. The blue region represents the "effective scatterer". $a_{ccw(cw)}$ and $b_{ccw(cw)}$ denote the complex-amplitude coefficients of the APMs outside and inside the "effective scatterer", respectively. (b1)-(b2) Definition of the effective reflection ($R_{cw}$) and transmission ($T$) coefficients of the APM at the "effective scatterer" for an incident CCW and CW APM, respectively. (c1)-(c2) Similar to (b1)-(b2) but for a complementary "effective scatterer". (d1)-(d2) Fundamental reflection ($\rho_j$) and transmission ($\tau_j$) coefficients of the APM at the nanohole $j$=1 and 2, respectively.

$$\Psi_{out}(r,\phi)=(a_{ccw}/v)\Psi_{ccw}(r,\phi)+a_{cw}\Psi_{cw}(r,\phi), \quad (1)$$

outside the "effective scatterer", and

$$\Psi_{in}(r,\phi)= b_{ccw}\Psi_{ccw}(r,\phi) +(b_{cw}/w)\Psi_{cw}(r,\phi), \quad (2)$$

inside the "effective scatterer", where $\Psi_{ccw}(r,\phi) =\Psi_{ccw}(r) \exp(ik_0 n_{eff} R_0 \phi)$ and $\Psi_{cw}(r,\phi)= \Psi_{cw}(r)\exp[ik_0 n_{eff} R_0 (2\pi-\phi)]$ with $\Psi=[\mathbf{E},\mathbf{H}]$ denote both the electric ($\mathbf{E}$) and the magnetic ($\mathbf{H}$) fields of the CCW and CW traveling APMs, respectively, $k_0=\omega/c$ (with $\omega$ and $c$ being the complex resonance angular frequency and the speed of light in the vacuum, respectively), and $n_{eff}$ is the complex effective index of the APM. $a_{ccw}/v$, $a_{cw}$, $b_{ccw}$, and $b_{cw}/w$ are the complex-amplitude coefficients of APMs normalized at $\phi=0$. Here $v=\exp(ik_0 n_{eff} R_0 \beta)$ and $w=\exp(ik_0 n_{eff} R_0 (2\pi-\beta))$ are the phase-shift factors of the APM traveling azimuthally over the range inside (from 0 to $\beta$) and outside (from $\beta$ to $2\pi$) the "effective scatterer", respectively.

Then the local chirality outside and inside the "effective scatterer" can be defined as [20]

$$\alpha_{out} = \frac{|a_{ccw}/v|^2 - |a_{cw}|^2}{|a_{ccw}/v|^2 + |a_{cw}|^2}, \quad \alpha_{in} = \frac{|b_{ccw}|^2 - |b_{cw}/w|^2}{|b_{ccw}|^2 + |b_{cw}/w|^2}, \quad (3)$$

respectively. The values of $\alpha_{out(in)}$ are within [-1,1], and $\alpha_{out(in)}=\pm 1$ implies a perfect chirality with a pure CCW (for "+") or CW (for "−") APM, respectively.

*Globally imperfect chirality at EPs.* —First, we study the impact of the nanohole size on the chirality $\alpha_{out(in)}$ at an EP. In the following calculation, the refractive indices of the microcavity, the nanoholes and the surrounding medium are set to be 2, 1 and 1 (air), respectively. Other parameters are $R_0$= 1.6μm, $d_1$ = 0.04 μm and $d_2$ = 0.0480 μm, with $r_2$ gradually increasing from $r_{2,0}$=0.108816 μm to $3r_{2,0}$, and $\theta_j$=2arcsin($r_j/4R_0$) ($j$=1, 2). Note that $r_2=r_{2,0}$

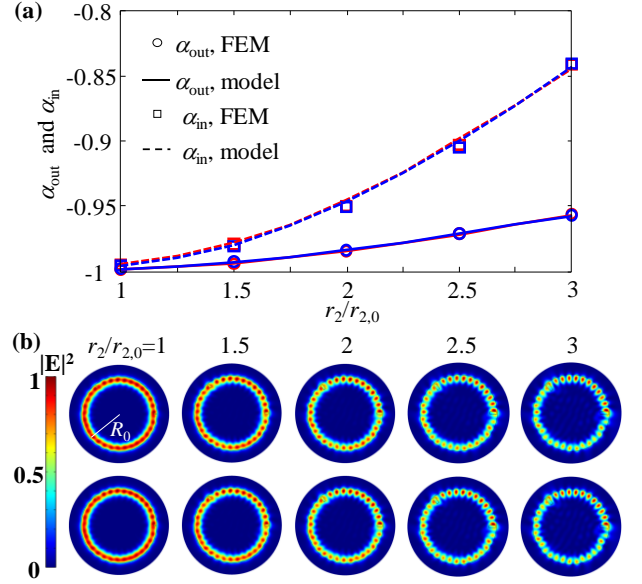

Figure 2 (a) Chiralities $\alpha_{out(in)}$ at the EPs plotted as functions of size $r_2$ of nanohole 2. The circles and squares show the FEM results of $\alpha_{out}$ and $\alpha_{in}$, respectively. The solid and dashed curves show the model prediction of $\alpha_{out}$ and $\alpha_{in}$, respectively. The blue and red curves show the results of the pair of nearly degenerate resonant modes. (b) Electric-field intensity distribution at the EPs obtained with the FEM shown in (a). The five columns are for different $r_2$, and the two rows are for the pairs of nearly degenerate resonant modes.

corresponds to the case of weak scattering, and the scattering effects of the nanohole become stronger with the increase of $r_2$ (see SM, Sec. 1.3 [28]). Here we consider the resonant modes corresponding to the unperturbed WGMs with electric vector along the $z$-direction, an azimuthal number $m$=16, a resonance wavelength around 1 μm, and a quality factor ($Q$) close to $10^5$.

By simultaneously scanning $r_1$ (around $r_2$) and $\beta$ (around 130°) (see SM, Sec. 2.3 [28]) with the other parameters fixed, an EP is found when the eigenfrequencies of a pair of resonant modes become degenerate and their corresponding waveform patterns become identical. The rigorous numerical results of $\alpha_{out(in)}$ can be obtained by extracting the APM coefficients with the mode-orthogonality theorem [33] from the electromagnetic field of resonant modes, which are solved with the full-wave finite-element method (FEM) performed by COMSOL MULTIPHYSICS [26,29].

Figure 2(a) shows that for the EP achieved with relatively weak scatterers, the chiralities outside (circles) and inside (squares) the "effective scatterer" are both almost perfect (at $r_2=r_{2,0}$ for instance, $\alpha_{out}$=−0.9987 and $\alpha_{in}$=−0.9949). Consequently, the electric-field intensities outside and inside the "effective scatterer" both exhibit a traveling wave pattern as shown in the first column in Fig. 2(b).

However, for the EPs achieved with strong scatterers (at $r_2=3r_{2,0}$ for instance), the chiralities outside and inside the "effective scatterer" both become imperfect, and the

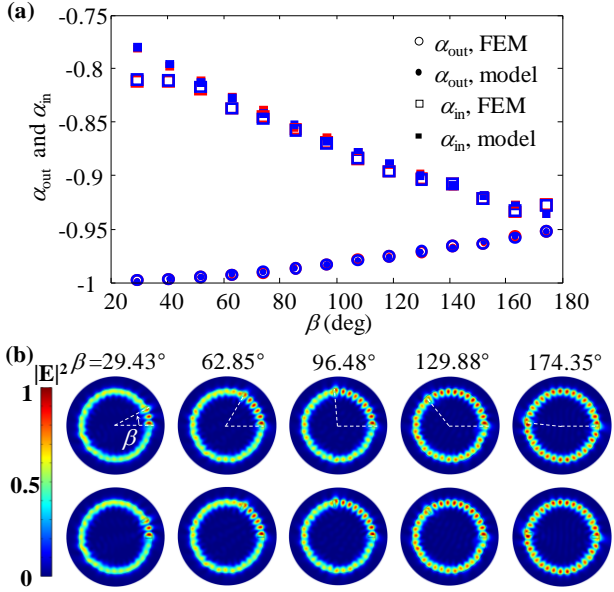

Figure 3 (a) Chiralities $\alpha_{\text{out(in)}}$ at EPs plotted as functions of $\beta$. The hollow (solid) circles and the squares denote the full-wave FEM results (model predictions) of $\alpha_{\text{out}}$ and $\alpha_{\text{in}}$, respectively. The blue and red colors correspond to two resonant modes. (b) Electric-field intensity distribution at the EPs for different $\beta$ values corresponding to the FEM results shown in (a). The different columns are for different $\beta$, and the two rows are for the pair of nearly degenerate resonant modes.

chirality within the larger azimuthal range divided by the two scatterers is stronger, i.e. $|\alpha_{\text{in}}|<|\alpha_{\text{out}}|<1$. Accordingly, the electric-field intensities exhibit a quasi-standing wave pattern formed by two counter-propagating APMs. This is surprising since EPs are always associated with a globally perfect chirality ($\alpha_{\text{out}}=\alpha_{\text{in}}=\pm 1$) in previous literatures [19-21,23].

*Locally perfect chirality at* EPs. —Second, we study the impact of the relative azimuthal angle $\beta$ between two strong scatterers on the chirality $\alpha_{\text{out(in)}}$ at EPs. The parameters (see SM, Sec. 2.3 [28]) in the following calculation are the same as those in Fig. 2, except that $r_2$ is fixed at a relatively large value of $2.5r_{2,0}$ to form a strong scattering.

Figure 3(a) shows that for the EPs achieved at large $\beta$, the chiralities outside (circles) and inside (squares) the "effective scatterer" are both imperfect, and there is $|\alpha_{\text{in}}|<|\alpha_{\text{out}}|<1$ similar to the case in Fig. 2(a). For instance, $\alpha_{\text{out}}=-0.9514$ and $\alpha_{\text{in}}=-0.9286$ for $\beta$ close to $174°$. Consequently, the electric-field intensities outside and inside the "effective scatterer" at the EPs do not possess an ideal traveling wave pattern, as shown in the last column in Fig. 3(b) (more results for different $\beta$ can be found in SM, Sec. 5 [28]).

However, for the EPs achieved at small $\beta$, Fig. 3(a) shows that $\alpha_{\text{out}}$ tends to be perfect ($|\alpha_{\text{out}}|\approx 1$), while $\alpha_{\text{in}}$ is weaker ($|\alpha_{\text{in}}|<1$), which implies a tendency of locally perfect chirality. For instance, for the EP with $\beta=29.43°$, Fig. 3(a) shows that $\alpha_{\text{out}}=-0.9978$ and $\alpha_{\text{in}}=-0.8087$.

Accordingly, the electric-field intensity outside the "effective scatterer" exhibits a distinct traveling wave pattern. In contrast, the field intensity inside the "effective scatterer" does not, as shown in the first column in Fig. 3(b). This locally perfect chirality is entirely different from the globally perfect chirality ($\alpha_{\text{out}}=\alpha_{\text{in}}=\pm 1$) for weak scatterers as shown in Fig. 2 and from the commonly reported results in previous literatures [19-21,23].

To understand the imperfect chirality at EPs for strong scatterers, we will apply the APM multiple-scattering model [26] to the present case of strong scatterers.

*Theoretical model*. —By considering a multiple-scattering process that incorporates the elastic transmission and reflection of APMs at the "effective scatterer", a set of coupled-APM equations can be written as [26]

$$a_{\text{ccw}} = a_{\text{ccw}} wT + a_{\text{cw}} wR_{\text{ccw}}, \quad (4a)$$

$$a_{\text{cw}} = a_{\text{ccw}} wR_{\text{cw}} + a_{\text{cw}} wT, \quad (4b)$$

where $T$, $R_{\text{cw}}$ and $R_{\text{ccw}}$ denote the effective scattering coefficients of the APMs at the "effective scatterer" as defined in Figs. 1(b1)-(b2). The effective scattering coefficients can be further derived from a Fabry-Pérot-like model [26],

$$R_{\text{cw(ccw)}} = \rho_{1(2)} + \frac{v^2 \rho_{2(1)} \tau_{1(2)}^2}{1 - v^2 \rho_1 \rho_2}, \quad T = \frac{v\tau_1\tau_2}{1 - v^2 \rho_1 \rho_2}, \quad (5)$$

where $\rho_j$ and $\tau_j$ ($j=1,2$) denote the fundamental reflection and transmission coefficients of the APM at nanohole $j$, respectively [as defined in Figs. 1(d1)-(d2)]. Note that $n_{\text{eff}}$, $\rho_j$ and $\tau_j$ all weakly depend on the frequency $\omega$ [29]. Therefore, the frequency dependence of $R_{\text{cw(ccw)}}$ and $T$ is dominated by the frequency dependence of $v=\exp(ik_0 n_{\text{eff}} R_0 \beta)$ which varies rapidly with the frequency $\omega$ via $k_0=\omega/c$.

The complex resonance frequencies (eigenvalues) of the two split modes, denoted by $\omega_+$ and $\omega_-$, can be obtained by solving the nontrivial solution of Eq. (4). By setting the determinant of the coefficient matrix of Eq. (4) to be zero, one can determine the $\omega_+$ and $\omega_-$ by solving two transcendental equations [26],

$$w(\omega_\pm) = \frac{1}{T(\omega_\pm) \pm R(\omega_\pm)}, \quad (6)$$

where $R(\omega)=\sqrt{R_{\text{cw}}(\omega)R_{\text{ccw}}(\omega)}$, and the $R(\omega_+)$ [respectively, $-R(\omega_-)$] denotes one of the two single-valued branches of $R(\omega)$ at $\omega=\omega_+$ (respectively, $\omega=\omega_-$). The single-valued branches of $R(\omega_+)$ and $-R(\omega_-)$ for $\omega_+\neq\omega_-$ have no specific relation for the present case of strong scatterers, which is different from $R(\omega_+)\approx R(\omega_-)$ for the case of weak scatterers [26].

Substituting Eq. (6) into Eq. (4), one can obtain the nontrivial solutions [26],

$$\frac{a_{\text{ccw}}(\omega_\pm)}{a_{\text{cw}}(\omega_\pm)} = \pm\sqrt{\frac{R_{\text{ccw}}(\omega_\pm)}{R_{\text{cw}}(\omega_\pm)}}, \quad (7)$$

where $a_{\text{ccw}}$ or $a_{\text{cw}}$ is determined by the normalization of the resonant mode. Besides the "effective scatterer" defined in [26], here we further define a complementary "effective scatterer" with an azimuthal range from $\beta$ to $2\pi$ as sketched in Figs. 1(c1)-(c2). The resultant merit is that $b_{\text{ccw}}/b_{\text{cw}}$ can be obtained in the same way as $a_{ccw}/a_{cw}$ with the following replacements in Eqs. (4)-(7), $a_{cw(ccw)} \to b_{cw(ccw)}$, $w \to v$, $v \to w$, $R_{cw(ccw)} \to R'_{cw(ccw)}$, $T \to T'$, $\rho_{1(2)} \to \rho_{2(1)}$, $\tau_{1(2)} \to \tau_{2(1)}$ (see SM, Secs. 2.1 and 2.2 [28]). This merit is crucial for explaining the chirality features at EPs over the whole azimuthal range of $[0,2\pi]$.

With the solved $a_{ccw}/a_{cw}$ and $b_{ccw}/b_{cw}$ inserted into Eq. (3), the chiralities outside and inside the "effective scatterer" can be expressed as,

$$\alpha_{out} \approx \frac{|R_{ccw}|-|R_{cw}|}{|R_{ccw}|+|R_{cw}|}, \quad \alpha_{in} \approx \frac{|R'_{ccw}|-|R'_{cw}|}{|R'_{ccw}|+|R'_{cw}|}, \quad (8)$$

respectively, where $|v| \approx |w| \approx 1$ in view of $\text{Im}(k_0 n_{eff} R_0) \approx 0$ [see Fig. 4(a)]. Equation (8) indicates that a perfect chirality $|\alpha_{out}|=1$ ($|\alpha_{in}|=1$) *is equivalent to* a unidirectional reflectionless APM at the (complementary) "effective scatterer", i.e. $R_{cw}=0$ or $R_{ccw}=0$ ($R'_{cw}=0$ or $R'_{ccw}=0$).

*Validity of the model.* —For the model predictions, the EPs are obtained by simultaneously scanning $r_1$ and $\beta$ with the other parameters being the same as the FEM results. As shown in Figs. 2 and 3, the chiralities predicted by the model agree well with those obtained with the full-wave FEM. More results to validate the model can be found in SM, Secs. 2.3 and 2.4 [28].

*Explanation of the conditional globally perfect chirality at EPs.* —The shifts of complex resonance frequencies of the two split resonant modes induced by the weak scatterers are quite small. It implies $k_0 n_{eff} R_0 \approx m$ [as confirmed numerically in Fig. 4(a)] and $\omega_\pm \approx \omega_- \approx \omega_0$ ($\omega_0$ being the complex resonance frequency for the unperturbed WGM).

Therefore, for the case of weak scatterers, the effective scattering coefficients $R_{cw(ccw)}$ and $T$ are approximately independent of the frequency $\omega$ due to $v \approx \exp(im\beta)$ and the weak dependence of $\rho_j$ and $\tau_j$ on $\omega$ [see Eq. (5)]. Then Eq. (6) becomes

$$w(\omega_\pm) = \frac{1}{T \pm R}, \quad (9)$$

where $R$ and $-R$ denote the two single-valued branches of $\sqrt{R_{cw} R_{ccw}}$ that is approximately independent of $\omega$. This implies that $R$ and $-R$ become related, which is different from the general situation in Eq. (6). When the system is steered to the EP, there is $w(\omega_+)=w(\omega_-)$ due to the degeneracy of the eigenfrequency ($\omega_+=\omega_-$), which yields $R=0$, i.e. $R$ at the square-root branch point [2,34]. Consequently, there is $R_{cw}=0$ or $R_{ccw}=0$ at the EP, which will result in a perfect chirality $|\alpha_{out}|=1$ according to Eq. (8).

Similarly, $R'_{cw(ccw)}$ and $T'$ are approximately independent of $\omega$ due to $w \approx \exp[im(2\pi-\beta)]$ for weak scatterers. Analogous to Eq. (9), one can obtain

$$v(\omega_\pm) = \frac{1}{T' \pm R'}, \quad (10)$$

by considering the complementary "effective scatterer", where $R' = \sqrt{R'_{cw} R'_{ccw}}$. Consequently, there is $R'=0$, i.e. $R'_{cw}=0$ or $R'_{ccw}=0$ at the EP, which causes a perfect chirality $|\alpha_{in}|=1$. Besides, it can be proved that $b_{cw(ccw)} \approx 0$ if $R_{cw(ccw)}=0$ for weak scatterers [26], which results in a

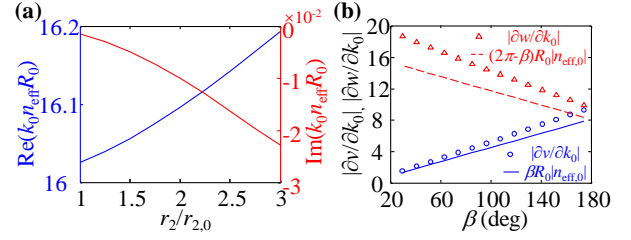

Figure 4 (a) Real and imaginary parts of $k_0 n_{eff} R_0$ plotted as functions of the size $r_2$ of nanohole 2, which are obtained with the model at the EPs already shown in Fig. 2. (b) $|\partial v/\partial k_0|$ (circles) and $|\partial w/\partial k_0|$ (triangles) at the EPs for the different $\beta$ values already shown in Fig. 3. The solid and dashed curves show the results of $\beta R_0 |n_{eff,0}|$ and $(2\pi-\beta) R_0 |n_{eff,0}|$, respectively.

globally perfect chirality ($\alpha_{out} \approx \alpha_{in} \approx \pm 1$) at the EP.

*Explanation of the conditional locally perfect chirality at EPs.*—For the case of strong scatterers, the frequency dependence of ($R_{cw(ccw)}$, $T$) and ($R'_{cw(ccw)}$, $T'$) (dominated by the frequency dependence of $v$ and $w$, respectively) could be remarkably different, which will result in the locally perfect chirality at EPs. This can be understood in view of

$$\left|\frac{\partial v}{\partial k_0}\right| = \beta R_0 \left|\frac{\partial (k_0 n_{eff})}{\partial k_0}\right| |v| \approx \beta R_0 |n_{eff}|, \quad (11)$$

and

$$\left|\frac{\partial w}{\partial k_0}\right| = (2\pi-\beta) R_0 \left|\frac{\partial (k_0 n_{eff})}{\partial k_0}\right| |w| \approx (2\pi-\beta) R_0 |n_{eff}|, \quad (12)$$

where $|\partial(k_0 n_{eff})/\partial k_0| \approx |n_{eff}|$ due to the weak dependence of $n_{eff}$ on $\omega$. Equation (11) indicates that $|\partial v/\partial k_0|$ tends to be 0 with the decrease of $\beta$, which means a weaker dependence of $R_{cw(ccw)}$ and $T$ on $\omega$. Consequently, Eq. (9) holds which leads to $|\alpha_{out}|=1$ for small $\beta$ as shown in Fig. 1(a). Differently, $|\partial w/\partial k_0|$ increases with the decrease of $\beta$, which means a strong dependence of $R'_{cw(ccw)}$ and $T'$ on $\omega$. Thus Eq. (10) does not hold which results in $|\alpha_{in}|<1$ for large $\beta$, as shown in Fig. 2(a). The validity of Eqs. (11) and (12) is confirmed in Fig. 4(b) where $n_{eff}$ approximately takes $n_{eff,0}=1.6186+9.9531 \times 10^{-6} i$ for the unperturbed WGM. The evolution of reflection coefficients $R_{cw(ccw)}$ and $R'_{cw(ccw)}$ near the EP can be found in SM, Sec. 4 [28].

This locally perfect chirality can be seen as reasonable in view that the "effective scatterer" is generally different from the complementary "effective scatterer" (with some exceptions, e.g. two identical scatterers with $\beta=\pi$), which will lead to different reflection coefficients ($R_{cw(ccw)} \neq R'_{cw(ccw)}$) of the APMs. Besides the above explanation, a logic crosscheck on the locally perfect chirality can be found in SM, Sec. 3 [28].

Following the above explanation, $|\alpha_{out}|>|\alpha_{in}|$ shown in Fig. 2(a) can be understood in view of the smaller relative azimuthal angle of the "effective scatterer" ($\beta \approx 130°$) than that of the complementary "effective scatterer" ($2\pi-\beta \approx 230°$).

*Discussion of previous works.* —In previous works [19-21,23], an effective Hamiltonian of the cavity-scatterer system is projected upon a global basis of the CW and

CCW traveling WGMs, which cannot describe the local chirality (as shown in Figs. 2 and 3). Furthermore, the eigenvalues and eigenvectors of the Hamiltonian are independent of the frequency (see Eq. (17) and (18) in Ref. [19]). As a result, a globally perfect chirality ($\alpha_{out}=\alpha_{in}=\pm 1$) at EPs is predicted, which holds approximately for weak scatterers but is not accurate for strong scatterers [27].

*Conclusions*. — Our study reveals that the chirality of the resonant modes at EPs of a WGM microcavity perturbed by two strong nanoscatterers is generally imperfect. The chirality will tend to be globally perfect for weak scatterers or to be locally perfect for small relative azimuthal angle between the two scatterers. With a first-principle-based model that incorporates an intuitive multiple-scattering process of the APMs, the conditional appearance of perfect chirality at EPs is attributed to the weak frequency-dependence of the effective APM scattering coefficients at the "effective scatterer", and correspondingly, the general imperfect chirality at EPs is due to the strong frequency-dependence of the effective APM scattering coefficients. Resultantly, a stronger chirality at EPs is observed in the larger azimuthal region divided by the two scatterers.

The discovered general features of the imperfect chirality at the EP will enrich the understanding of EP, not only in optics but also in other systems, such as microwave [16], acoustic [35] and quantum systems [36]. The previous criterion of EP in experiments [8,20,25], i.e. a perfect chirality or the equivalent unidirectional no-reflection [see Eq. (8)], should be refreshed for strong scatterers. Besides, the proposed peculiar features of the chirality at EP may promote unconventional applications in the on-chip chiral photonics, for instance, a switchable unidirectional and bidirectional lasing in different azimuthal ranges of the microcavity based on the locally perfect chirality at the EP.

Financial support from the National Key Research and Development Program of China (2019YFA0705000); National Natural Science Foundation of China (11734009, 62075104, 12004197, 12034010, 12074199, 12134007, 92050111, 92050114); 111 Project (B07013).

*liuht@nankai.edu.cn
†bofang@nankai.edu.cn
‡yang@seas.wustl.edu
§zhanggq@nankai.edu.cn
¶jjxu@nankai.edu.cn

# SUPPLEMENTAL MATERIAL
# Imperfect chirality at exceptional point in optical whispering-gallery microcavities

The Supplemental Material provides some details for the dependence of the scattering coefficients on the location, refractive index and size of the nanoscatterer, additional logic and numerical test on the validity of the model, some additional explanation of the conditional perfect chirality at exceptional points (EPs), evolution of reflection coefficients as functions of the relative azimuthal angle $\beta$ between the two nanoholes, and additional results of the electric-field intensities at the EPs for different $\beta$.

CONTENT
1. Dependence of the scattering coefficients on the location, refractive index and size of the nanoscatterer
    1.1 Dependence of the scattering coefficients on the location of the nanoscatterer
    1.2 Dependence of the scattering coefficients on the refractive index of the nanoscatterer
    1.3 Dependence of the scattering coefficients on the size of the nanoscatterer
2. Additional logic and numerical test on the validity of the model
    2.1 Theoretical model for the complementary "effective scatterer"
    2.2 Equivalence between model equations for the "effective scatterer" and the complementary "effective scatterer"
    2.3 Validity of the model in predicting the structural parameters at the EPs
    2.4 Validity of the model in predicting the evolution of complex resonance frequencies near the EPs
3. A logic crosscheck on the locally perfect chirality at the EP for strong scatterers
4. Evolution of reflection coefficients as functions of the relative azimuthal angle between the two nanoholes
5. Additional results of the electric-field intensities at the EPs for different relative azimuthal angle between the two nanoholes
References

## 1. DEPENDENCE OF THE SCATTERING COEFFICIENTS ON THE LOCATION, REFRACTIVE INDEX AND SIZE OF THE NANOSCATTERER

### 1.1 Dependence of the scattering coefficients on the location of the nanoscatterer

In this subsection, we provide the numerical results to show the dependence of the scattering coefficients on the location of the nanoscatterer.

As shown in Fig. S1, the reflection ($\rho$) and transmission ($\tau$) coefficients of the azimuthally propagating mode (APM) at an internal (solid curves) or external nanoscatterer (dashed curves) are plotted as a function of the gap $d$ between the nanoscatterer and the rim of the microcavity [as sketched in the insets in Fig. S1(a)]. Here we consider the scattering coefficients of the fundmental APM that will form the unperturbed transverse-magnetic (TM, electric vector along the invariant $z$-direction) WGMs with a high quality factor ($Q$) close to $10^5$, which are the APMs considered in the model. The normalized electric field $|\mathbf{E}|$ of the fundamental APM with wavelength setting to be 1 μm are shown in the insets in Fig. S1(a) (the solid curves). The size of the internal and external nanoscatterers are fixed, and takes the value of the largest nanohole 2 ($r_{2,0}=3r_{2,0}=0.32645$ μm) as shown in Fig. 2(a) in the main text. The refractive indices are setted to be $n_c=n_{ex}=2$, $n_{sur}=n_{in}=1$ (air), where $n_c$, $n_{sur}$, $n_{ex}$ and $n_{in}$ denote the refractive indices of the microcavity, the surrounding medium, the external and internal nanoscatterer, respectively.

In Fig. S1, it is seen that for both the internal and external nanoscatterers, $|\rho|$ (red curves) increase and $|\tau|$ (blue curves) decrease with the decrease of $d$, implying a stronger scattering effect of the APM at the nanoscatterer closer to the rim of the microcavity. Besides, $|\rho|$ (respectively, $|\tau|$) for the internal nanoscatterer (solid curves) is larger (respectively, smaller) than that for the external nanoscatterer (dashed curves), implying a stronger scattering effect of the APM at the internal nanoscatterer than that at the external nanoscatterer. Thus, for the case of nanoholes located far away from the rim of microcavity, e.g. $d \approx 0.3\mu m$ in Ref. [1], or the case of the nanoparticles located in the evanescent field outside the microcavity [2,3], the scatterering effect could be not strong enough to observe a distinct imperfect chirality at the EP.

The dependence of the scattering coefficients on the location of the nanoscatterer can be understood in view that the scattered electromagnetic field $[\mathbf{E}^{(s)}, \mathbf{H}^{(s)}]$ satisfies the following Maxwell's equations [4],

$$\nabla \times \mathbf{E}^{(s)} = i\omega\mu\mathbf{H}^{(s)}, \nabla \times \mathbf{H}^{(s)} = -i\omega\varepsilon\mathbf{E}^{(s)} - i\omega\Delta\varepsilon\mathbf{E}^{(i)}, \tag{S1}$$

where $\mathbf{E}^{(i)}$ represents the incident electric field (i.e. the fundamental APM), $\varepsilon$ and $\mu$ are the permittivity and permeability of the system in the presence of a nonmagnetic scatterer, respectively. $\Delta\varepsilon$ denotes the permittivity change between the systems in the presence and absence of the scatterer. There are $\Delta\varepsilon = n_{in}^2 - n_c^2 = 1-2^2 = -3$ and $\Delta\varepsilon = n_{ex}^2 - n_{sur}^2 = 2^2-1 = 3$ within the region of the internal and external nanoscatterer, respectively. Equation (S1) indicates that the scattered field can be regarded as the field radiated in the presence of the scatterer by an effective current source $-i\omega\Delta\varepsilon\mathbf{E}^{(i)}$. So the scattered electric field $\mathbf{E}^{(s)}$ will be stronger if the incident electric field $\mathbf{E}^{(i)}$ within the region of the scatterer (with its size and $|\Delta\varepsilon|$ fixed) is stronger. As shown in the inset in Fig. S1(a), the incident electric field $\mathbf{E}^{(i)}$ within the region of the internal scatterer is much stronger than that within the region of the external scatterer, which explains the stronger scattering effect of the APM at an internal nanoscatterer than that at an external nanoscatterer.

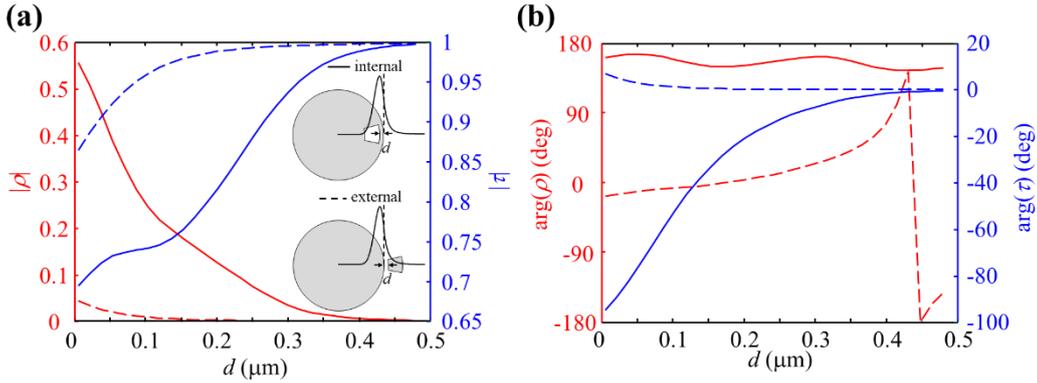

Figure S1. Modulus (a) and argument (b) of the reflection coefficient $\rho$ (red curves) and transmission coefficient $\tau$ (blue curves) of the APM at an internal (solid curves) or external nanoscatterer (dashed curves) plotted as a function of the gap $d$ between the nanoscatterer and the rim of the microcavity. The insets show the schematics of the two kinds of nanoscatterer and the normalized electric field $|\mathbf{E}|$ of the fundamental APM.

## 1.2 Dependence of the scattering coefficients on the refractive index of the nanoscatterer

In this subsection, we provide the numerical results to show the dependence of the scattering coefficients on the refractive index of the nanoscatterer.

As shown in Fig. S2, the reflection ($\rho$) and transmission ($\tau$) coefficients of the APM at the internal (solid curves) or external nanoscatterer (dashed curves) are plotted as a function of $\Delta\varepsilon$. Here both $n_{in}$ and $n_{ex}$ take the values ranging from 1 to 3.5. The size of the internal and external nanoscatterers are fixed and takes the value of the smallest nanohole 2 ($r_{2,0}=0.108816$ μm) as shown in Fig. 2(a) in the main text, and the gap $d$ is fixed to be $d=d_2=0.0480$ μm. It is seen that $|\rho|$ increases and $|\tau|$ decreases with the increase of $|\Delta\varepsilon|$, implying a stronger scattering effect at the nanoscatterer with higher refractive-index contrast. This can be understood in view that according to Eq. (S1), the scattered electric field $\mathbf{E}^{(s)}$ will be stronger for larger $|\Delta\varepsilon|$ when the incident field and the size of the scatterer are fixed. Thus, filling the nanoholes with a material of high refractive index to form a stronger scatterer is a potential way to realize an imperfect chirality at the EP.

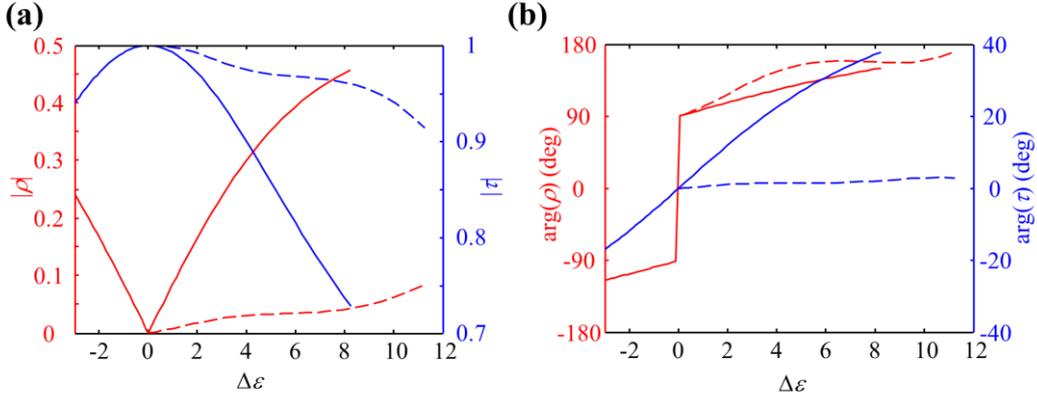

Figure S2. Modulus (a) and argument (b) of the reflection coefficient $\rho$ (red curves) and transmission coefficient $\tau$ (blue curves) of the APM at the internal (solid curves) or external nanoscatterer (dashed curves) plotted as a function of $\Delta\varepsilon$.

## 1.3 Dependence of the scattering coefficients on the size of the nanoscatterer

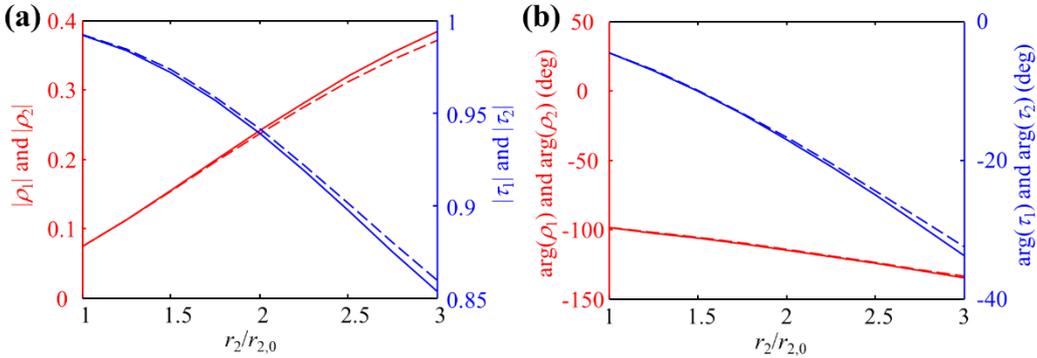

Figure S3. Modulus (a) and argument (b) of the reflection coefficient $\rho_j$ (red curves) and transmission coefficient $\tau_j$ (blue curves) of the APM at the single nanohole 1 ($j=1$, solid curves) or 2 ($j=2$, dashed curves) plotted as functions of the size $r_2$ of nanohole 2. The size $r_1$ of nanohole 1 takes the values at the EPs as shown in Fig. S4.

In this subsection, we provide the numerical results to show the dependence of the scattering coefficients on the size of the nanoscatterer.

As shown in Fig. S3, the reflection ($\rho_j$) and transmission ($\tau_j$) coefficients of the APM at the single naonohole 1 or 2 ($j$=1 or 2) are plotted as a function of size $r_2$ of nanohole 2. The sizes of naonoholes 1 and 2 take the values at the EPs as shown in Fig. 2(a) in the main text, and the values of the size $r_1$ of nanohole 1 can be found in Fig. S4 in subsection 2.3. It is seen that $|\rho_j|$ ($j$=1, 2) increases and $|\tau_j|$ decreases with the increase of $r_j$. So one can tune the scattering effect by changing the size of the nanoholes as done in Fig. 2 in the main text.

## 2. ADDITIONAL LOGIC AND NUMERICAL TEST ON THE VALIDITY OF THE MODEL

### 2.1 Theoretical model for the complementary "effective scatterer"

In this subsection, we will provide the theoretical model for the complementary "effective scatterer" defined with an azimuthal range from $\beta$ to $2\pi$, as illustrated in Figs. 1(c1) and (c2) by the blue region. Fully parallel to Eq. (4) in the main text, a set of coupled-APM equations can be written as,

$$b_{ccw} = b_{ccw} v T' + b_{cw} v R'_{ccw}, \tag{S2a}$$

$$b_{cw} = b_{ccw} v R'_{cw} + b_{cw} v T', \tag{S2b}$$

where $b_{ccw}$ and $b_{cw}$ denote the complex-amplitude coefficients of the CCW and CW APMs outside the complementary "effective scatterer", respectively, and $v=\exp(ik_0 n_{eff} R_0 \beta)$ is the phase-shift factor of the APM traveling azimuthally over the azimuthal range outside (from 0 to $\beta$) the complementary "effective scatterer". $T'$, $R'_{cw}$ and $R'_{ccw}$ denote the effective transmission and reflection coefficients of the APMs at the complementary "effective scatterer". The coupled-APM equation [Eq. (S2a)] can be understood intuitively in view that the CCW APM with coefficient $b_{ccw}$ results from two contributions: one from the transmission ($T'$) of itself (with coefficient $b_{ccw}$ and a phase-shift factor $v$) at the complementary "effective scatterer", and the other from the reflection ($R'_{ccw}$) of the CW APM (with coefficient $b_{cw}$ and the same phase-shift factor $v$) at the complementary "effective scatterer". Equation (S2b) can be understood in a similar way.

The effective scattering coefficients can be further expressed as

$$R'_{cw(ccw)} = \rho_{2(1)} + \frac{w^2 \rho_{1(2)} \tau^2_{2(1)}}{1 - w^2 \rho_1 \rho_2}, \tag{S3a}$$

$$T' = \frac{w \tau_1 \tau_2}{1 - w^2 \rho_1 \rho_2}. \tag{S3b}$$

By solving Eq. (S2) in a way fully parallel to Eq. (4), one can determine the $\omega_+$ and $\omega_-$ by solving two transcendental equations

$$v(\omega_\pm) = \frac{1}{T'(\omega_\pm) \pm R'(\omega_\pm)}, \tag{S4}$$

where $R'(\omega) = \sqrt{R'_{cw}(\omega) R'_{ccw}(\omega)}$. The associated eigenvectors are determined by the ratios

$$\frac{b_{ccw}(\omega_\pm)}{b_{cw}(\omega_\pm)} = \pm \sqrt{\frac{R'_{ccw}(\omega_\pm)}{R'_{cw}(\omega_\pm)}}. \tag{S5}$$

## 2.2 Equivalence between model equations for the "effective scatterer" and the complementary "effective scatterer"

In this subsection, we will provide a logic crosscheck on the equivalence between the coupled-APM equation (4) in the main text and equation (S2), which are written for the "effective scatterer" [as sketched in Fig. 1(b)] and the complementary "effective scatterer" [Fig. 1(c)], respectively. In Ref. [2], it has been proved that Eq. (4) can be derived from the model equations [i.e. the following Eq. (S6)] without using the concept of the "effective scatterer". In the following, we will prove that Eq. (S2) can be derived as well from Eq. (S6) in a way parallel to Eq. (4), which then yields the logic equivalence between Eq. (4) and (S2).

By considering a dynamical multiple-scattering process of the clockwise (CW) and counter clockwise (CCW) APMs at the two individual nanoscatterers, the coupled-APM equations without using the concept of the "effective scatterer" can be written as [2],

$$a_{ccw} = a_{cw} w \rho_2 + b_{ccw} v \tau_2, \tag{S6a}$$

$$a_{cw} = a_{ccw} w \rho_1 + b_{cw} v \tau_1, \tag{S6b}$$

$$b_{ccw} = b_{cw} v \rho_1 + a_{ccw} w \tau_1, \tag{S6c}$$

$$b_{cw} = a_{cw} w \tau_2 + b_{ccw} v \rho_2. \tag{S6d}$$

From Eqs. (S6c)-(S6d), one can obtain,

$$a_{ccw} = \frac{b_{ccw} - b_{cw} v \rho_1}{w \tau_1}, \tag{S7a}$$

$$a_{cw} = \frac{b_{cw} - b_{ccw} v \rho_2}{w \tau_2}. \tag{S7b}$$

Substituting Eq. (S7) into Eqs. (S6a)-(S6b) and after simple operations, one can obtain,

$$b_{ccw} \tau_2 = b_{cw} v \rho_1 \tau_2 + b_{ccw} w v \tau_1 \left( \tau_2^2 - \rho_2^2 \right) + b_{cw} w \rho_2 \tau_1, \tag{S8a}$$

$$b_{cw} \tau_1 = b_{ccw} w \rho_1 \tau_2 + b_{cw} w v \tau_2 \left( \tau_1^2 - \rho_1^2 \right) + b_{ccw} v \rho_2 \tau_1. \tag{S8b}$$

Substituting the expression of $b_{cw} \tau_1$ [the right side of Eq. (S8b)] into the third term in the right side of Eq. (S8a) and after some operations, one can obtain,

$$b_{ccw} = b_{ccw} v \left( \frac{w \tau_1 \tau_2}{1 - w^2 \rho_1 \rho_2} \right) + b_{cw} v \left( \rho_1 + \frac{w^2 \rho_2 \tau_1^2}{1 - w^2 \rho_1 \rho_2} \right), \tag{S9}$$

Similarly, by substituting the expression of $b_{ccw} \tau_2$ [the right side of Eqs. (S8a)] into the first term in the right side of Eq. (S8b), one can obtain,

$$b_{cw} = b_{ccw} v \left( \rho_2 + \frac{w^2 \rho_1 \tau_2^2}{1 - w^2 \rho_1 \rho_2} \right) + b_{cw} v \left( \frac{w \tau_1 \tau_2}{1 - w^2 \rho_1 \rho_2} \right). \tag{S10}$$

It is seen that Eqs. (S9) and (S10) simply become Eq. (S2) with the coefficients $T'$, $R'_{cw}$ and $R'_{ccw}$ defined by Eq. (S3).

## 2.3 Validity of the model in predicting the structural parameters at the EPs

In this subsection, we will check the validity of the model in predicting the structural parameters $r_1$ and $\beta$ at the EPs against the full-wave finite-element method (FEM) results. Note that the sectorial nanohole is chosen here to facilitate the calculation, and the shape of the nanohole is unrestricted, as explained in the SM in Ref. [2]. For the model prediction, the values of $r_1$ and $\beta$ at EP can be determined by seeking $r_1$ and $\beta$ to satisfy $\omega_+(r_1,\beta)=\omega_-(r_1,\beta)$ since the solved $\omega_+$ and $\omega_-$ are functions of the two real-valued structural parameters $r_1$ and $\beta$, i.e. $\omega_\pm=\omega_\pm(r_1,\beta)$.

For the EPs already shown in Fig. 2 in the main text, Fig. S4 shows that the $r_1$ and $\beta$ predicted by the model (solid curves) agree well with those by the FEM (circles).

For the EPs shown in Fig. 3, Fig. S5 also shows good agreement between the $r_1$ and $\beta$ predicted by the model (dots) and those by the FEM results (circles). Here note that at the EPs, $r_1$ only changes moderately with a quasi-periodic increase of $\beta$ near $(n+1/2)\pi/m$ (with $m=16$ and $n$ being an integer). This can be understood in view of the approximate invariance of $r_1$ along with the periodicity of $\beta$ at the EPs for weak scatterers [2]. The slight deviations of the model from the full-wave FEM results are due to the presence of the residual field other than the APMs considered in the model [2]. Note that for the case of $\beta>\pi$ (correspondingly, $n>15$), the EPs with a positive chirality ($\alpha_{out}>0$, $\alpha_{in}>0$) can be obtained in view of the symmetry between $\beta>\pi$ and $\beta<\pi$.

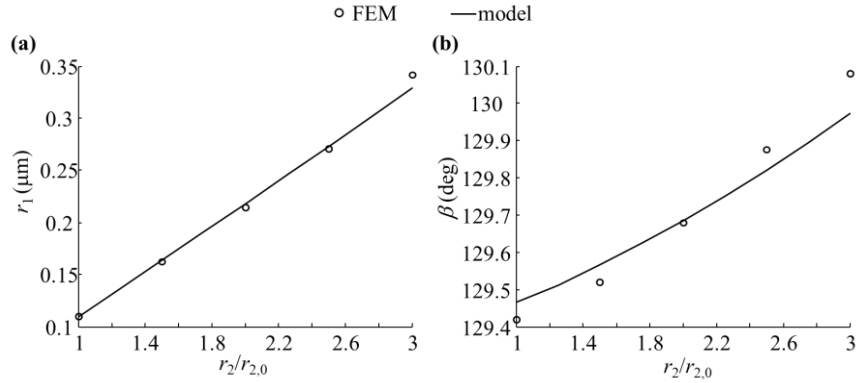

Figure S4. Model predictions (solid curves) and the full-wave FEM results (circles) of $r_1$ (a) and $\beta$ (b) at the EPs already shown in Fig. 2.

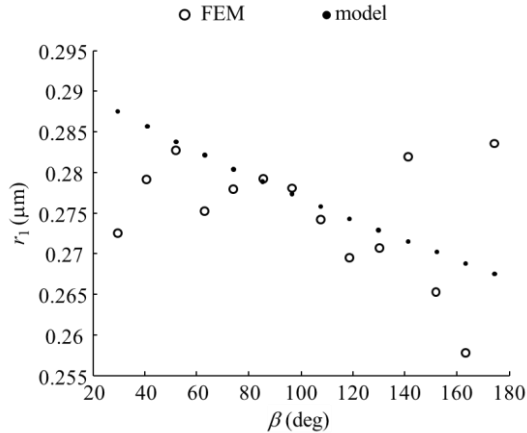

Figure S5. Model predictions (dots) and the FEM results (circles) of $r_1$ and $\beta$ at the EPs already shown in Fig. 3.

## 2.4 Validity of the model in predicting the evolution of complex resonance frequencies near the EPs

In this subsection, we will check the validity of the model in predicting the evolution of the complex resonance frequencies as functions of $\beta$ near EPs.

As shown in Fig. S6, the dimensionless complex resonance frequency $\Omega_c=\omega_c R_0/c$ is plotted as a function of $\beta$. The EP (at $\beta=62.85°$ already shown in Fig. 3) predicted by the model and that by the FEM are indicated by the green solid and dashed arrows in the insets of Fig. S6, respectively. The blue and red curves correspond to the pair of split modes with complex resonance frequencies $\omega_c=\omega_+$ and $\omega_-$ [see Eqs. (6) and (S4)], respectively. It is seen that the $\Omega_c$ predicted by the model (solid curves) agrees well with that by the full-wave FEM (dashed curves).

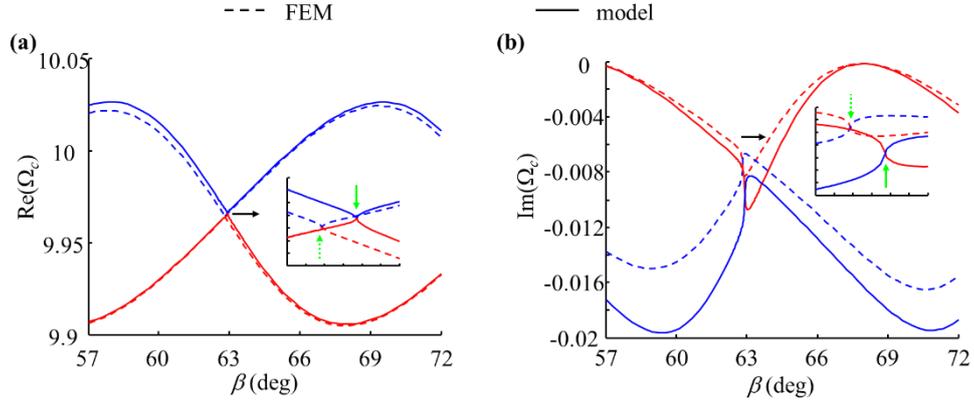

Figure S6. Real (a) and imaginary (b) part of the dimensionless complex resonance frequency $\Omega_c=\omega_c R_0/c$ plotted as a function of $\beta$ around the EP with $\beta=62.85°$ already shown in Fig. 3. The dashed and solid curves show the full-wave FEM results and the model predictions, respectively. The blue and red curves are for the pair of split modes with complex resonance frequencies $\omega_c=\omega_+$ and $\omega_-$, respectively. The insets show magnifications of the region $\beta\in[62.74°, 63.08°]$ near the EP.

## 3. A LOGIC CROSSCHECK ON THE LOCALLY PERFECT CHIRALITY AT EPs FOR STRONG SCATTERERS

Besides the explanation in the main text, here we will provide a logic crosscheck on the locally perfect chirality at EPs shown in Fig. 3 for strong scatterers. For this purpose, we will prove that the chiralities inside and outside the "effective scatterer" cannot be perfect simultaneously at the EP for strong scatterers and small $\beta$.

First, we assume that the chirality outside the "effective scatterer" is perfect at the EP, i.e. $|\alpha_{out}|=1$. Then we can obtain $R_{cw}=0$ or $R_{ccw}=0$ according to Eq. (8) in the main text. Then Eq. (6) reduces to $w=1/T$. With Eq. (5) inserted into, $w=1/T$ becomes

$$w = \frac{1-v^2\rho_1\rho_2}{v\tau_1\tau_2} \Leftrightarrow v = \frac{1-v^2\rho_1\rho_2}{w\tau_1\tau_2}. \tag{S13}$$

Equation (S13) can be rewritten as

$$v = \frac{1-w^2\rho_1\rho_2}{w\tau_1\tau_2} + \frac{(w^2-v^2)\rho_1\rho_2}{w\tau_1\tau_2} = \frac{1}{T'} + C_1. \tag{S14}$$

where the second equality is obtained with the use of Eq. (S3b) and $C_1$ is defined as the second term on the right side of the first equality. The right side of Eq. (S14) can be further rewritten as

$$v = \frac{1}{T' + C_2}, \quad (S15)$$

where $C_2 = -C_1 T'^2/(1+C_1 T') = -\rho_1\rho_2(w^2-v^2)T'^2/[\tau_1\tau_2 w + \rho_1\rho_2(w^2-v^2)T']$, which is nonzero due to the strong scatterers ($\rho_1\rho_2 \neq 0$ and $T' \neq 0$) and small $\beta$ (in general $w \neq v$). Comparing Eq. (S15) and Eq. (S4), one can obtain $\pm R' = C_2 \neq 0$, i.e. $R'_{cw} \neq 0$ and $R'_{ccw} \neq 0$, which implies an imperfect chirality $|\alpha_{in}| < 1$ according to Eq. (8).

## 4. EVOLUTION OF REFLECTION COEFFICIENTS AS FUNCTIONS OF RELATIVE AZIMUTHAL ANGLE BETWEEN THE TWO NANOHOLES

It is mentioned in the main text that a strong or weak frequency-dependence of the effective scattering coefficients will result in worse or better chirality at the EP, respectively [as explained with Eqs. (9) and (10)]. To further confirm it, in this section, we will provide the model prediction of the evolution of the reflection coefficients $R(\omega_+)$, $-R(\omega_-)$, $R'(\omega_+)$ and $-R'(\omega_-)$ [defined after Eqs. (6) and (S4)] as functions of the relative azimuthal angle $\beta$ between the two nanoholes. Here we take the EP with $\beta=62.85°$ (already shown in Fig. S6 and Fig. 3 in the main text) as an example, and scan $\beta$ around the EP.

As confirmed in Fig. S7, the difference between $R'(\omega_+)$ and $R'(\omega_-)$ is larger than that between $R(\omega_+)$ and $R(\omega_-)$ for a fixed $\beta$ away from the EP. Besides, the value of $|R(\omega_\pm)|$ ($3\times10^{-3}$) is much smaller than that of $|R'(\omega_\pm)|$ ($1.5\times10^{-2}$) at the EP, where $R(\omega_+)=-R(\omega_-)$ and $R'(\omega_+)=-R'(\omega_-)$ in view of $\omega_+=\omega_-$ and Eqs. (6) and (S4). These different behaviors between $R(\omega)$ and $R'(\omega)$ are due to the smaller azimuthal range of the "effective scatterer" ($\beta=62.85°$) than that of the complementary "effective scatterer" ($2\pi-\beta=297.15°$), which then results in a stronger frequency-dependence of $R'(\omega)$ than that of $R(\omega)$ [see Eqs. (11) and (12)].

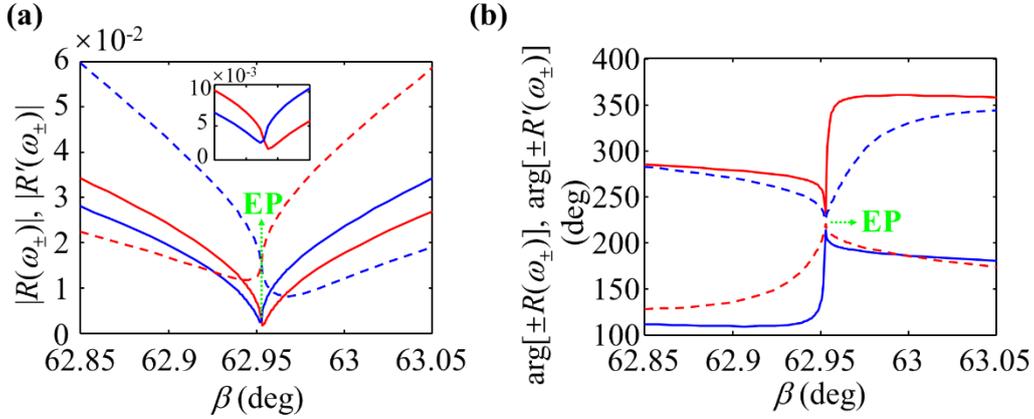

Figure S7. Evolution of the modulus (a) and argument (b) of $R(\omega_+)$ (blue solid curves), $-R(\omega_-)$ (red solid curves), $R'(\omega_+)$ (blue dashed curves) and $-R'(\omega_-)$ (red dashed curves) as functions of $\beta$ around the EP [indicated by the green arrows, which is the EP with $\beta=62.85°$ shown in Figs. 3(a) and S6]. The inset in (a) shows a magnification of the region $\beta\in[62.945°, 62.96°]$ near the EP.

## 5. ADDITIONAL RESULTS OF THE ELECTRIC-FIELD INTENSITIES AT THE EPs FOR DIFFERENT RELATIVE AZIMUTHAL ANGLE BETWEEN THE TWO NANOHOLES

In this section, we provide additional results of the electric-field intensities at the EPs shown in Fig. 3 in the main text. Figures S8(a1)-(a14) and (b1)-(b14) provide the results for the pair of nearly degenerate resonant modes at the EPs, and different columns correspond to different relative azimuthal angle $\beta$ between the two scatterers of nanoholes. It is seen that for the EPs achieved at small $\beta$, the electric-field intensity outside the "effective scatterer" exhibits a distinct travelling wave pattern, which however is not true for the field inside the "effective scatterer". While for the EPs achieved at large $\beta$, neither of the electric-field intensities outside and inside the "effective scatterer" possesses an ideal travelling wave pattern.

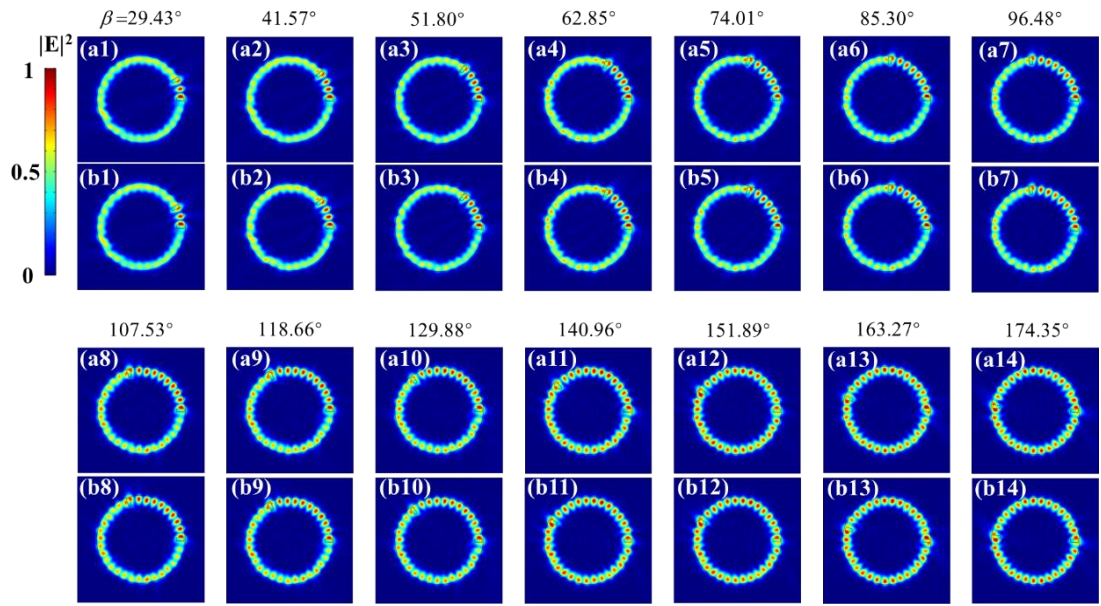

Figure S8. Electric-field intensities $|\mathbf{E}|^2$ (normalized with a maximum of 1) at the EPs obtained with the FEM shown in Fig. 3 in the main text. The different columns are for different $\beta$, and (a1)-(a14) and (b1)-(b14) are for the pair of nearly degenerate resonant modes.